# Modified Bose-Einstein and Fermi-Dirac statistics if excitations are localized on an intermediate length scale: Application to non-Debye specific heat

## Ralph V. Chamberlin* and Bryce F. Davis

Department of Physics, Arizona State University, Tempe, AZ 85287-1504, USA

**Abstract**: Disordered systems show deviations from the standard Debye theory of specific heat at low temperatures. These deviations are often attributed to two-level systems of uncertain origin. We find that a source of excess specific heat comes from correlations between quanta of energy if phonon-like excitations are localized on an intermediate length scale. We use simulations of a simplified Creutz model for a system of Ising-like spins coupled to a thermal bath of Einstein-like oscillators. One feature of this model is that energy is quantized in both the system and its bath, ensuring conservation of energy at every step. Another feature is that the exact entropies of both the system and its bath are known at every step, so that their temperatures can be determined independently. We find that there is a mismatch in canonical temperature between the system and its bath. In addition to the usual finite-size effects in the Bose-Einstein and Fermi-Dirac distributions, if excitations in the heat bath are localized on an intermediate length scale, this mismatch is independent of system size up to at least $10^6$ particles. We use a model for correlations between quanta of energy to adjust the statistical distributions and yield a thermodynamically consistent temperature. The model includes a chemical potential for units of energy, as is often used for other types of particles that are quantized and conserved. Experimental evidence for this model comes from its ability to characterize the excess specific heat of imperfect crystals at low temperatures.

*e-mail:  ralph.chamberlin@asu.edu

**I. Introduction**

Disordered materials show "universal" features in their low-temperature thermal properties that are not fully understood [1-3]. One such feature is a peak in the excess specific heat (above that of the Debye model) at around 10 K. Most crystals show a similar peak in specific heat, but with amplitudes that decrease with decreasing disorder [4-7]. The standard Debye model gives the specific heat of plane-wave vibrations (phonons) in ideal crystals, but disordered systems show an excess density of states in the vicinity of the Boson peak that contributes to the specific heat at low temperatures [8,9]. A model based on Einstein oscillators localized to each site has been used to predict the minimum thermal conductivity of highly-disordered systems at higher temperatures [10]. At lower temperatures the properties are often attributed to an ensemble of two-level systems [11], with phonon scattering on a relatively narrow range of intermediate length (IL) scales [12-15]. Here we use a simplified model that combines these features: a two-level system of Ising-like "spins" (binary degrees of freedom) with Einstein-like oscillators ("demons") that scatter on IL scales. It is based on the Creutz model [16], with the advantage that the system and its bath have identical energy spacing, thereby facilitating conservation of total energy at every step. Here we use a simplified Creutz model, with the additional advantage that the entropies of both the system and its bath can be calculated independently, yielding two distinct canonical temperatures at every step.

The zeroth law of thermodynamics states that when two systems are in thermal equilibrium they have the same temperature, $T$, but the law does not specify the statistical distribution needed to calculate $T$. Although many quantized systems can be characterized by Bose-Einstein (BE) or Fermi-Dirac (FD) statistics, these distributions rely on the usual assumptions of the Boltzmann factor [17,18]: the system must be able to borrow an unlimited amount of energy from an effectively infinite heat bath, the thermal contact between the system and its bath must be weak but much faster than the microscopic dynamics, the heat bath must contain a smooth distribution of closely-spaced energies, and the quanta of energy must be uncorrelated. In other words, adding one unit of energy to the system must not change the probability of adding a second unit. Here we explore some consequences of using an explicit heat bath, with quantized energies for the system and bath, so that the local energy is exactly conserved at every step. We find that assuming the BE and FD distributions yields unequal canonical temperatures for the system and its bath. One source of this mismatch in canonical temperature comes from finite-size effects, as found in simulations and theory of critical clusters during nucleation [19,20]. Indeed, for random-walk (RW) demon motion we find that the mismatch in temperature decreases with increasing system size. However, an unanticipated result is that the magnitude of the temperature mismatch is independent of system size for IL dynamics, where most excitations travel freely for a number of lattice sites before being reflected. We model this persistent mismatch in canonical temperature using an effective interaction between neighboring quanta of energy, combined with a chemical potential to accommodate conservation of energy. Thus this temperature mismatch can be attributed to correlations between units of energy when energy is quantized and conserved. We show that this model adjusts the BE and FD distributions, and yields a single $T$. Then we show that the model provides good agreement with measured excess specific heat from imperfect crystals at low $T$.

    This paper is organized as follows. In the Background section we describe the simplified Creutz model and give an exact expression for its total entropy. Then we approximate this entropy by assuming large systems, and use the usual definition of temperature to derive the BE and FD distributions. In the Simulation Details section we describe the system of spins, the heat

bath of demons, and the initialization procedure for the simulations. Then we specify the steps in the simplified Creutz model that reflect excited demons on an IL scale. In the Results section we first establish that the thermal equilibrium of the simplified Creutz model is governed by maximizing the sum of exact entropies from the system and its bath, not the usual canonical-ensemble approximations, which explains the finite-size effects found for RW demon motion. Then we show that significant deviations from BE and FD statistics occur independent of system size if energetic demons reflect on an IL scale. Next we describe an adsorption model for energy quanta that modifies the BE and FD distributions, and gives an offset for their average energies that yields thermodynamically-consistent expressions for temperature. Finally we show that this energy-quanta adsorption model also gives good agreement with the excess specific heat measured in imperfect crystals. Our Conclusions are that 1) correlations between quanta of energy in a simplified model cause deviations from the Boltzmann factor that modify the BE and FD distributions, 2) the excess entropy from these correlations may provide an explanation for the non-Debye specific heat in disordered materials, and 3) a similar breakdown in canonical-ensemble statistics may occur in many real systems if the heat bath contains localized excitations that are quantized.

**II. Background**

Computer simulations have shown that a nonlinear correction to the Boltzmann factor improves agreement between the standard Ising model and the measured response of many materials [18-22]. Here we use a related model, introduced by Creutz [16], with two types of particles: Ising-like "spins" that have fixed positions on a lattice (the system), and kinetic-energy-carrying "demons" that move through the lattice (the heat bath). Unlike the Ising model in the canonical ensemble, where spins may spontaneously increase their potential energy with probability proportional to the Boltzmann factor, the Creutz model has an explicit heat bath of Einstein-like oscillators (the demons). The demons act as a source of kinetic energy, simulating a microcanonical ensemble where the total energy is exactly conserved at every step. The dynamics of the Creutz model involves choosing a demon, moving it to a neighboring site in a chosen direction, and testing the potential energy of the spin at that site for possible inversion. If the potential energy decreases or stays the same the spin always inverts, giving any extra energy to the demon. If the potential energy tries to increase, the demon must have enough energy to invert the spin without reaching negative kinetic energy. Thus, the standard Creutz model is a type of cellular automata, with dynamics that can be made deterministic [23]. Here we investigate a simplified Creutz model, where the interaction between spins is replaced by a uniform interaction with an external field ($B$). Key advantages of the simplified Creutz model are that the exact canonical distributions and entropies are known for both the system and its bath.

Canonical temperature is found from the change in entropy with respect to energy using the fundamental equation of thermodynamics: $1/T = \Delta S/\Delta \mathcal{H}$. Note that differences are used instead of differentials because energy is quantized. The Hamiltonian is $\mathcal{H} = \varepsilon m + \varepsilon n$, where $m$ and $n$ are the number of energy quanta in the spins and demons, respectively, and $\varepsilon = 2B$ is the unit of energy. We use Boltzmann's expression for statistical entropy $S = k_B \ln(W)$, where $k_B$ is Boltzmann's constant and $W$ is the multiplicity of the microstate, which is the number of microstates that yield a single macrostate. This $S$ has the advantage that it is defined for every microstate, with no need for time- or ensemble-averaging [24]. In the simplified Creutz model, the entropy depends on the number of demons ($N$) and spins ($M$), and the number of energy units in each. For non-interacting binary spins the multiplicity is the binomial coefficient: $W_m = M! /$

$[m!(M–m)!]$, whereas for Einstein oscillators [25]: $W_n = (n+N–1)! / [n!(N–1)!]$. Assuming that entropy is additive for the system and its bath, the total entropy is $S_{m,n}/k_B = \ln(W_m) + \ln(W_n)$. For a fixed number of spins and demons, changes in $S_{m,n}$ arise when quanta of energy are transferred between the system and its bath, $\Delta m = -\Delta n$. The most accurate simple expression for discrete differences [26] gives $\Delta S = (S_{m+½,n} – S_{m–½,n})\Delta m + (S_{m,n+½} – S_{m,n–½})\Delta n$, or

$$\frac{\Delta S}{k_B} = \ln\frac{M-m+½}{m+½}\Delta m + \ln\frac{n+N-½}{n+½}\Delta n \qquad (1)$$

Note that using differences instead of derivatives also improves accuracy by avoiding Stirling's approximation for the factorials.

If the ½-terms are neglected Eq. (1) yields two distinct canonical temperatures: a demon temperature $T_D$ and a spin temperature $T_S$. Specifically, $T_D$ is obtained using $\Delta S/\Delta \mathcal{H}$ with $\Delta\mathcal{H}=\varepsilon \Delta n$ for the change in demon energy, then assuming $n \gg ½$ gives $\varepsilon/k_B T_D = \ln[(n+N)/n]$. Note that this $T_D$ is the temperature usually used for Creutz-model simulations. Exponentiating the logarithm, with $T_D \rightarrow T$, yields the BE distribution $n_0/N = 1/(e^{\varepsilon/k_B T} - 1)$. A chemical potential is not needed here because the number of demons is fixed. Similarly, $T_S$ is obtained using $\Delta S/\Delta \mathcal{H}$ with $\Delta\mathcal{H} = \varepsilon \Delta m$ for the change in spin energy, then assuming $m \gg ½$ gives $\varepsilon/k_B T_S = \ln[(M–m)/m]$. Inverting this equation with $T_S \rightarrow T$ yields the FD distribution $m_0/M = 1/(e^{\varepsilon/k_B T} + 1)$. From computer simulations we find that $T_D$ and $T_S$ often differ significantly.

**III. Simulation details**

We simulate simple-cubic lattices of $M$ non-interacting binary spins in an external field $B=1$, with a thermal bath of $N$ demons. The total number of lattice sites is set to a value of $M=2^3$ to $96^3$, with an equal number of demons per site, $N=M$. In the usual Ising model, the spin at site $i$ may be up ($\sigma_i=1$) or down ($\sigma_i=-1$), with a potential energy that depends on the relative alignment of nearest-neighbor spins. For the simplified Creutz model the potential energy comes only from the external field, giving the unit of energy $\varepsilon = 2$ in the Hamiltonian $\mathcal{H} = \varepsilon m + \varepsilon n$. Here the number of potential-energy quanta (from spins aligned opposite to the field) is $m = \sum_{i=1}^{M}(1-\sigma_i)/2$, and the number of kinetic-energy quanta is $n = \sum_{j=1}^{N} k_j$, where $k_j$ is a non-negative integer giving the number of units of energy in the $j^{th}$ demon. Note that the units of potential energy at each site (0 or 1) simulates the statistics for the number of fermions in a given state, whereas the units of kinetic energy on each demon ($k_j = 0,1,2,3…$) simulates the statistics of bosons in a given state. The total quanta of energy per site [energy density $\eta = \mathcal{H}/(M\varepsilon)$] is set to a constant value $0.0005 \leq \eta \leq 2.5$. Energy is initialized with all spins up (minimum potential energy) by adding one unit of energy to a randomly chosen demon until the energy density equals or exceeds the set value $\eta$. Thus, because energy is quantized, the exact energy density may differ from $\eta$, especially at low temperatures on small lattices. However, if energy is initialized using a consistent procedure, the difference between the set value and actual $\eta$ is reproducible. Furthermore, we do not include simulations on the smallest systems at lowest temperatures where the set value and actual $\eta$ differed by more than 10% (note the missing data in Fig. 1 for small lattices at low energies). Simulations start with a random velocity direction for each demon: $\hat{v}_j = \pm\hat{x}$, $\pm\hat{y}$, or $\pm\hat{z}$. Demon dynamics involves choosing a demon at random, moving it one lattice site in its direction $\hat{v}_j$, then attempting to flip the spin at the new site. If the spin is up ($\sigma_i = 1$) it flips only if the demon has positive energy ($k_j > 0$), which decreases the demon energy by one unit. Whereas if the spin is down ($\sigma_i = -1$) the spin always flips, which increases the demon energy by one unit. The average probability for demons of energy $k$ is given by: $\overline{p_k} = \sum_j \delta_{k,k_j} / \sum_{k,j} \delta_{k,k_j}$. Here $j$ is the sum over all steps and $\delta$ is the Kronecker delta. Thus each step adds one to every denominator, with one added to the $k^{th}$ numerator only if the demon has $k$ units of energy at the end of the step.

Demons move between neighboring sites in the lattice, with periodic boundary conditions on all external surfaces. For RW demon motion, the direction of each step is chosen using a random-number generator. For IL dynamics, each demon moves in its direction $\hat{v}_j$ unless reflected at an interface between regions. The regions are defined by subdividing the lattice into cube-shaped blocks with sides of length $L$, so that each region contains $L^3$ sites. If all demons reflect equally, there is negligible change in the equilibrium thermal properties. However, if the reflection probability is energy dependent, consistent with experiments [12-15] and theory [27], significant changes occur. Here we focus on reflections of excited demons, where demons with non-zero energy are reflected with probability $(1-\tau)$, so that they are transmitted through the interface with probability $\tau$. If a demon were to revisit the same site immediately after reflection, its energy transfer would be nullified, tending to trap energy at interfaces. Therefore, all reflections involve moving the demon back one site from which it came and inverting its velocity to $-\hat{v}_j$, so that energetic demons spend less time at interface sites. Each demon moves in only one-dimension (two possible directions), but because other demons move in the other dimensions, energy is transferred between all demons and all sites. Although this energy transfer can be quite slow for $\tau \ll 1$ where excitations are strongly localized, we ensure equilibrium behavior by verifying that there is negligible change in the average properties if the number of steps is reduced by an order of magnitude.

We start all simulations with no reflections, using $\tau=(1/2)^g$ with the power $g=0$ so that $\tau=1$. Then $\tau$ is decreased by increasing $g$ in integer steps. We find that the offset in canonical temperature increases with decreasing $\tau$ (increasing localization) until $\tau \sim 0.001$, where the properties become relatively stable. We obtain equilibrium values for IL dynamics by averaging six sets of simulations with $\tau \sim 0.001$; specifically $g=9, 10, 11, 12, 13$ and $14$. We find that there is negligible difference between the thermal-average properties for RW dynamics and long-range ballistic motion ($\tau=1$), where all demons maintain a constant velocity across the entire sample with periodic boundary conditions and no reflections. Thus, relative values for IL dynamics are obtained by subtracting the $g=0$ behavior. Simulations were made for $2 \times 10^4$-$2 \times 10^6$ Monte-Carlo sweeps for each value of $g$. For example, $2 \times 10^4$ sweeps were used for $M=96^3$ and $2 \times 10^6$ sweeps for $M \leq 24^3$. Thermal averages were taken from the final 60% of the sweeps. Thus, averages over $g=9$-$14$ for lattices containing $24^3$ sites come from $10^{10}$ Monte Carlo steps.

**IV. Results**

Figure 1 shows the difference in inverse canonical temperature as a function of the total energy density, $\eta = (n + m)/M$. The lower sets of symbols are from RW demon dynamics, with $\varepsilon/k_B T_S - \varepsilon/k_B T_D$ multiplied by the lattice size $M = 2^3$ to $16^3$. Hence this difference goes to zero in the thermodynamic limit, $M \rightarrow \infty$, yielding $T_D \rightarrow T_S \rightarrow T_{RW}$. A convenient expression for this RW canonical temperature comes from combining the BE and FD distributions with conservation of energy: $\varepsilon/k_B T_{RW} = \ln(1 - \eta + \sqrt{1 + \eta^2}) - \ln(1 + \eta - \sqrt{1 + \eta^2})$. However, the upper sets of symbols have $\varepsilon/k_B T_S - \varepsilon/k_B T_D$ multiplied by the length scale for demon localization, $L=3$-$12$, showing a significant mismatch in temperature that is independent of total system size, at least for $M = 6^3$ to $96^3$.

Finite-size effects in the canonical temperatures for RW demon motion can be explained by including the ½-terms in Eq. (1). Rewriting Eq. (1) by separating the canonical-ensemble and finite-size terms gives

$$\frac{\Delta S}{k_B} = \left[\frac{\varepsilon}{k_B T_S} + \ln\frac{1+\frac{1/2}{M-m}}{1+\frac{1/2}{m}}\right]\Delta m + \left[\frac{\varepsilon}{k_B T_D} + \ln\frac{1-\frac{1/2}{n+N}}{1+\frac{1/2}{n}}\right]\Delta n \qquad (2)$$

Entropy is maximized when energy is conserved ($\Delta n = -\Delta m$) if the quantities in square brackets in Eq. (2) are equated. Indeed, for RW motion $\varepsilon/k_B T_S - \varepsilon/k_B T_D$ closely follows the difference between the two logarithmic terms in Eq. (2), as shown by the dashed line in Fig. 1. This quantitative agreement with no adjustable parameters demonstrates that thermal equilibrium of the Creutz model is governed by maximizing the total entropy of the system plus bath, not the usual canonical-ensemble assumptions. Furthermore, for RW motion, finite-size corrections to the BE and FD distributions come from the logarithmic terms in Eq. (2). For IL dynamics, however, such mathematical corrections do not capture all features in the mismatch. Instead, the solid lines in Fig. 1 show that the mismatch can be characterized by an energy-quanta adsorption model, as described below.

Figure 2 is a semi-logarithmic plot of the average probability $\overline{p_k}$ of finding demons containing $k$ units of energy as a function of $k$. All simulation are on a medium-sized lattice, $M=24^3$. Solid lines from RW motion show that $\overline{p_k}$ decreases exponentially with increasing $k$, consistent with the Boltzmann factor. The constant magnitude of the inverse slope of these lines yields a single canonical temperature over at least nine orders of magnitude in $\overline{p_k}$. However, the dashed and dotted lines from IL dynamics show an energy-dependent slope. Indeed, close inspection reveals that compared to RW dynamics, IL dynamics yields: slightly higher $\overline{p_k}$ for $k$=0, slightly lower $\overline{p_k}$ for small $k$, and much higher $\overline{p_k}$ for large $k$. Thus there is an effective interaction that tends to increase the energy gap between the ground state and first excited state of the demons. The large increase in $\overline{p_k}$ at high $k$ is due primarily to longer lifetimes of excited demons that are localized by reflections. In any case, if energetic demons are localized on an IL scale, the bath of Einstein oscillators has an energy distribution that deviates from BE statistics and requires a nonlinear correction to the Boltzmann factor [18-22].

The symbols in Fig. 3 come from IL dynamics where energetic demons reflect at boundaries between regions with sides of length $L$=3 or 4, as given in the figure. RW dynamics (not shown in Fig. 3) would yield zero offset as $M\to\infty$, corresponding to the dot-dashed line. We emphasize that the maximum deviation from canonical statistics occurs for IL dynamics with $L$=3, returning to the canonical distributions for short-ranged RW or for long-ranged ballistic motion. This IL scale, larger than atomic spacing but smaller than sample size, may be related to the "universal" mean-free path of acoustic waves in disordered materials [14], and could yield an intrinsic coarse-graining length to facilitate entropy increase [28]. Open triangles show the offset in demon energy $(n-n_0)/N$ as a function of $k_B T_D/\varepsilon$, whereas closed triangles show the offset in spin energy $(m-m_0)/M$ as a function of $k_B T_S/\varepsilon$. Note the conspicuous increase in $(n-n_0)/N$ and shift to higher $k_B T_D/\varepsilon$, while $(m-m_0)/M$ decreases an equal amount (to conserve energy) and shifts to lower $k_B T_S/\varepsilon$. Dotted lines connect symbols from the same simulation at the same total energy. Thus, the horizontal offset between each pair of connected symbols shows the mismatch in canonical temperature, which exceeds 40% for $L$=3 at the highest energy.

The solid and dashed lines in Fig. 3 come from the BE and FD distributions using $T_D$. Specifically, the upper two solid lines show the BE distribution as a function of $T_D$, $1/(e^{\varepsilon/k_B T_D}-1) - n_0/N$, which agrees with simulations of $(n-n_0)/N$ exactly by definition of $T_D$. The other two solid lines show that the FD distribution as a function of $T_D$, $1/(e^{\varepsilon/k_B T_D}+1) - m_0/M$, has a small increase in energy, opposite to the decrease required by conservation of energy as shown by the simulations. The dashed lines that roughly follow the lower two sets of symbols again use $T_D$ in the FD function, but with the numerator and energy spacing adjusted to best fit the data, then plotted as a function of $T_S$. For $L=3$, the fit parameter $\varepsilon=1.93\pm0.02$ differs from the actual energy spacing $\varepsilon=2$, while the numerator $0.835\pm0.005$ also differs significantly from its actual value of 1. Similarly, some values of this FD function differ from the simulation values of $m/M$ by more than 10 standard deviations. Thus, for IL dynamics, the energies cannot be described by the BE and FD distributions alone.

To characterize deviations from the canonical distributions we use an interacting lattice-gas model for adsorption of particles on neighboring sites [29], adapted to treat quanta of energy. We find that the offset in canonical temperatures can be modeled by an effective potential $w<0$, which causes a correlation between energy quanta on neighboring sites. Although agreement with simulations may be better if all possible arrangements of occupied sites in a region are included, here for simplicity we consider only nearest-neighbor pairs. The partition function is: $\xi = 1 + 2ae^{-\varepsilon/k_B T_{RW}} + a^2 e^{-2\varepsilon/k_B T_{RW}} e^{-w/k_B T_{RW}}$, where $a=e^{\mu/k_B T_{RW}}$ is the energy-quantum activity, with $\mu$ the chemical potential. The three terms in $\xi$ come from: no energy, one unit of energy on either site, and one unit of energy on both sites, which includes the effective interaction potential. The probability of finding no energy on either site is $1/\xi$.

We find that this $1/\xi$ often favors a net transfer of energy from spins to demons. Thus, this model for correlations between neighboring quanta of energy gives an effective interaction that tends to align non-interacting spins. A possible explanation is that if neither site has energy, demons with zero energy pass through with no chance of acquiring energy, and hence no chance of being localized. In other words, if two neighboring spins are condensed into their ground state, they act as a relatively rigid pair that is less likely to localize demons. The increased entropy from delocalizing ground-state demons tends to increase the energy spacing to the first excited state, even though there is no direct interaction. This effective interaction can be thought of a type of entropic force favoring low-energy states when the spin-alignment bath has low entropy, similar to how low temperatures favor low-energy states when the heat bath has low entropy [22]. Using conservation of total energy, the density of energy in demons and spins becomes $n/N = n_0/N + A/\xi$ and $m/M = m_0/M - A/\xi$, respectively. Here we use $A$ as an adjustable parameter that governs the strength of the energy transfer mechanism. We also adjust $a$ to the best value for each set of simulations, as given in Table 1. If $w$ is released as an adjustable parameter, from twelve sets of simulations we find $w/\varepsilon = -3.99 \pm 0.03$. Thus, we use $w/\varepsilon = -4$ for accurate and reliable fits with only two adjustable parameter. From 9 sets of data we find $a = 0.442 \pm 0.018$ for localization lengths of $L = 4, 6,$ and $8$, but the value for $L=3$ is significantly different $a = 0.629 \pm 0.013$. A consistent equilibrium temperature is obtained by inverting either energy density, with the appropriate offset: $\varepsilon/k_B T = \ln[1/(n/N - A/\xi) + 1] \approx \ln[1/(m/N + A/\xi) - 1]$. The open circles in Fig. 3, with solid circles near their centers, show that the modified BE and FD distributions give good agreement with the energies of the demons and spins, and equilibrium temperatures that match.

Heat capacity in the Debye model is proportional to the size of the lattice, so that at low temperature it can be written as $C_D = DMT^3$. Here $D$ is a constant that depends on the number of degrees of freedom per lattice site. To characterize measured excess specific heat we assume that any extra energy adsorbed on a lattice site simply adds to its degrees of freedom, so that $D \rightarrow D + C(\bar{\varepsilon}/\varepsilon)(m - m_0)/(MA)$. Here $C$ is an excess specific heat constant, $(m - m_0)$ gives the additional units of energy at each site, $\bar{\varepsilon} = k_B T^2 \partial \ln(\xi)/\partial T$ is the average energy from the adsorption model, and the ratio $\bar{\varepsilon}/\varepsilon$ may be a simplistic way to count the extra units of energy when $\varepsilon$ comes from a distribution. The modified and normalized specific heat can be written as:

$$C_a/(MT^3) = D - C[2ae^{-\varepsilon/k_B T} + (2 + w/\varepsilon)a^2 e^{-2\varepsilon/k_B T} e^{-w/k_B T}]/\xi^2 \tag{3}$$

Note that because we find $w/\varepsilon < -2$, the quantity in square brackets is often negative, yielding excess specific heat for $C > 0$.

Symbols in Fig. 4 show the measured excess specific heat reported in the literature for three different crystalline substances. Solid curves are the best fits to these data using Eq. (3), yielding: $w/\varepsilon = -2.92 \pm 0.08$ and $a = 0.476 \pm 0.03$ (see Table 1). Note that $a$ is within experimental error of the value obtained from the simulations for $L \geq 4$, but that the magnitude of $w/\varepsilon$ indicates a 30% weaker interaction between neighboring quanta of energy in these real materials. Despite the simplistic way that Einstein-like oscillators are added to Debye-like vibrations, Eq. (3) gives good agreement with the excess specific heat in these samples. In the Creutz model, equal energy spacing simplifies the simulations and facilitates conservation of energy. From the measured excess specific heat of imperfect crystals, agreement with the model using equal energy spacing of the system and its bath could indicate a type of resonance, or other inherent connection between localized phonons and the two-level systems [30].

Debye theory describes the low-temperature specific heat of ideal crystals, which may require isotopic purity [5-7] to show no deviations from $T^3$ behavior. At the other extreme, amorphous materials exhibit a larger and broader excess peak, suggesting a range of interactions between a distribution of energy quanta. At higher temperatures, several experimental techniques have shown that the primary response of many disordered solids and supercooled liquids comes from independently relaxing regions [31-35], where excitations are localized on the length scale of nanometers. Thus, a full description of the thermal and dynamic properties of these substances may require models [36-40] and simulations [18-22] based on non-canonical distributions and nonlinear corrections to the Boltzmann factor, consistent with the results presented here.

**V. Conclusions**

We have shown that a simplified model of Ising-like spins coupled to a heat bath of Einstein-like oscillators exhibits significant deviations from the canonical Fermi-Dirac and Bose-Einstein distributions if excitations are localized on an intermediate length scale. The mechanism involves a breakdown in the assumptions of the Boltzmann factor due to the large energy-spacing and small size of each local thermal bath. We characterize the deviations using a chemical potential and effective interaction between units of energy, as is often done for other types of particles that are quantized and conserved. This effective interaction from the correlation between units of energy tends to increase the energy gap between the ground state and excited states of the Einstein oscillators. The energy-quanta adsorption model also gives good agreement with the measured excess specific heat that causes deviations from the Debye model in imperfect crystals at low temperatures, as shown in Fig. 4. Similar "universal" deviations from the Debye model are found in amorphous systems over a similar temperature range. Thus, correlations between neighboring units of energy due to a quantized thermal bath that is localized on an intermediate length scale may provide an explanation for the excess specific heat found in most materials at low temperatures.

For equilibrium simulations of the simplified Creutz model, Fig. 1 shows that there are significant differences between the canonical temperatures of a system and its bath. Although seeming to violate the zeroth law of thermodynamics, in fact this mismatch is due to deviations from the canonical ensemble that yield nonlinear corrections to the Boltzmann factor, as shown in Fig. 2. Figure 3 shows that a thermodynamically consistent temperature can be calculated by including correlations between units of energy. Figure 3 also shows that this energy-quanta correlation model adjusts the Fermi-Dirac and Bose-Einstein distributions to match the thermal equilibrium of a system of Ising-like spins coupled to a bath of Einstein-like oscillators when excitations are localized on an intermediate length scale. We speculate that similar adjustments to the canonical-ensemble distributions may be necessary for any real system if the thermal bath contains localized excitations that are quantized.

## VI. Acknowledgments

We thank N. Bernhoeft, V. Mujica, N. Newman, and G.H. Wolf for several helpful suggestions. We acknowledge technical assistance from the A2C2 computing facility. We thank the ARO for supporting this research (W911NF-11-1-0419).

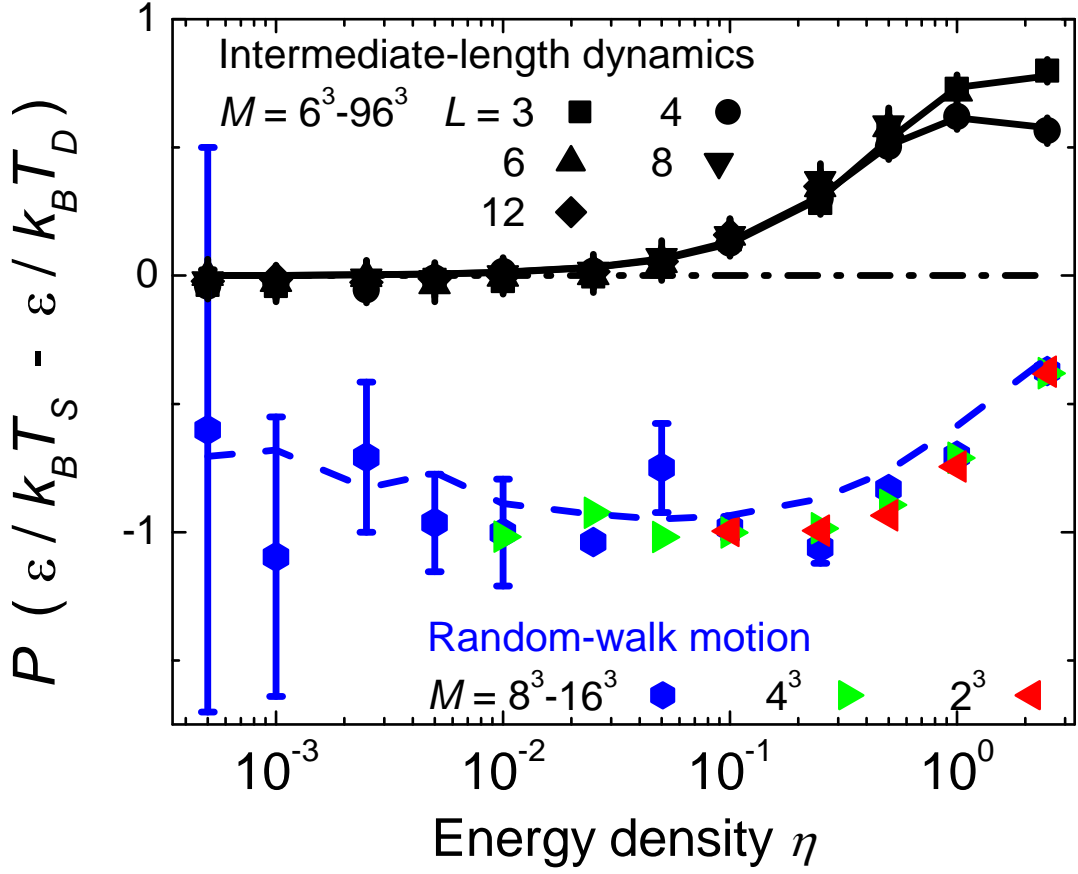

FIG. 1. (Color online) Difference between inverse canonical temperature of spins and demons as a function of the average energy per lattice site. The dot-dashed line at zero shows the behavior that would come from canonical-ensemble statistics. Symbol shape identifies the simulation size ($M$, given in the figure) and type of demon dynamics. Error bars (from averaging multiple simulation sizes) are visible if larger than the symbol size. The three lower sets of data come from RW demon motion, with the differences multiplied by the prefactor $P=M$, with $M = 2^3$ to $16^3$ lattice sites. Thus, this mismatch in canonical temperature goes to zero in the thermodynamic limit $M \rightarrow \infty$. However, for the five upper sets of data the prefactor is $P = L$, with localization lengths of $L=3$ to 12 sites inside systems of $M = 6^3$ to $96^3$ sites. Thus, for IL dynamics the mismatch in canonical temperature is independent of simulation size. The dashed line shows finite-sized deviations from BE and FD statistics if entropy is maximized using differences instead of derivatives, from the logarithmic terms in Eq. (2). Solid lines come from an energy-quanta adsorption model, as described in the text.

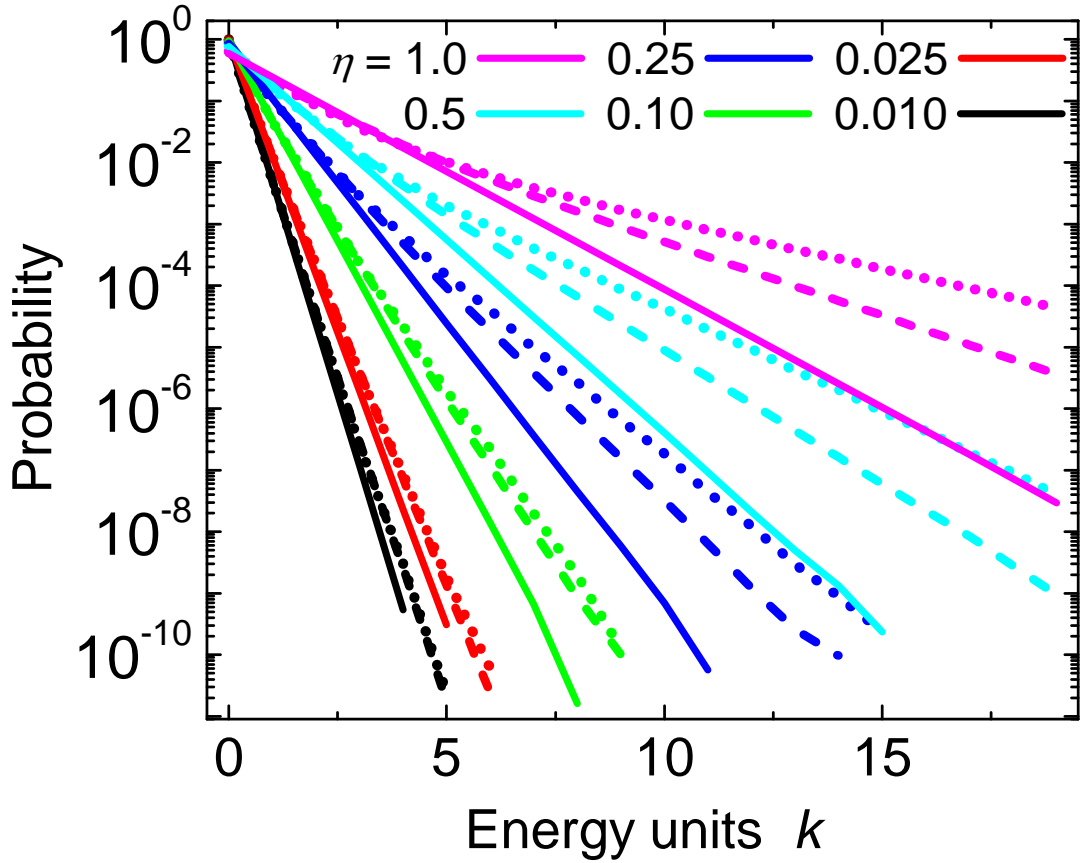

FIG. 2. (Color online) Average probability $\overline{p_k}$ of demon energy as a function of the number of energy quanta in the demon. Line style identifies the type of demon dynamics: RW (solid), IL dynamics in regions with sides of length $L=4$ (dashed) or $L=3$ (dotted). Different sets of lines come from different average energy densities, $\eta$, given in the figure. For RW dynamics, the constant slope of each solid line yields the same temperature for all energies. For IL dynamics, the dashed and dotted lines show energy-dependent slopes, indicating deviations from the Boltzmann factor. Clearly the probability of adding one unit of energy is not the same as adding a second unit. Indeed, the probability of finding some high-energy demons can be several orders of magnitude higher if they are localized on small length scales.

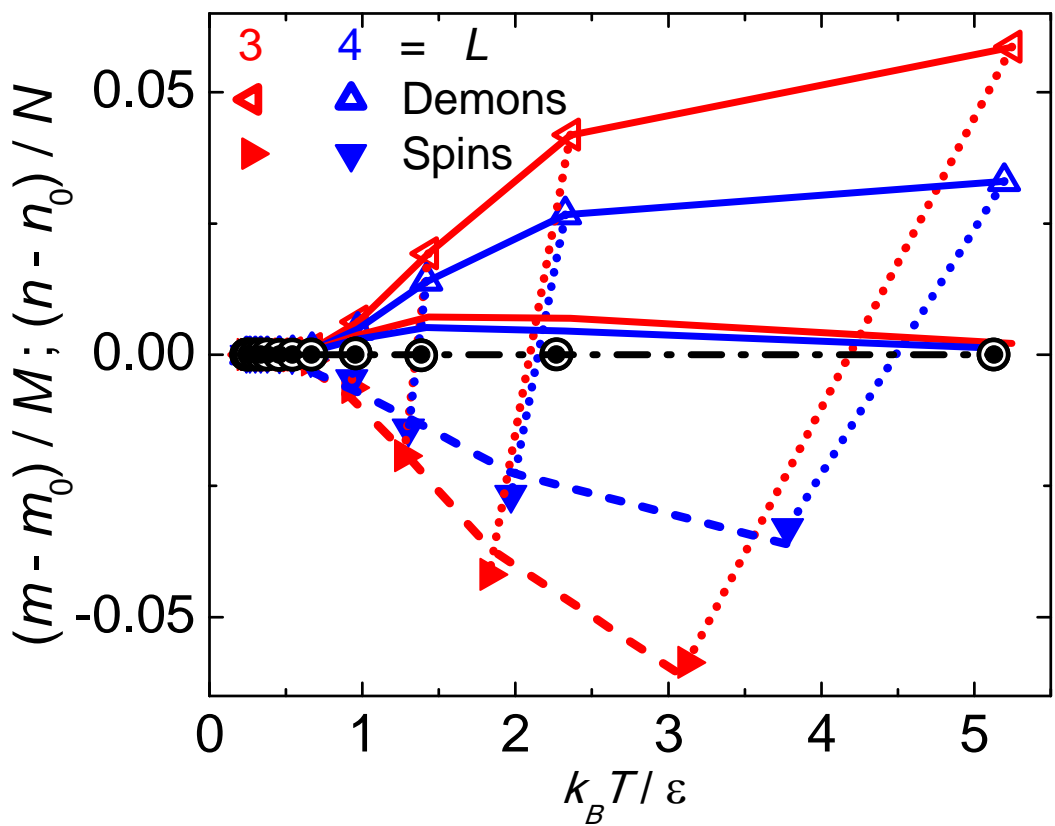

FIG. 3. (Color online) Difference between theoretical and simulated energy densities for IL demon dynamics. The differences are plotted as a function of canonical temperature (triangles) or corrected temperature (circles). Open and solid symbols identify the behavior of demons and spins, respectively. Symbol shape also identifies the localization length for demon motion, $L$, as given in the figure. Dotted lines connect pairs of points from the same simulation at the same total energy. Note that the finite slope of these lines shows the mismatch in canonical temperature, which exceeds 40% for $L=3$ at the highest energy. Also note that the increase in demon energy above the center line matches the decrease in spin energy below the line, as required by conservation of energy. Other lines come from various models, as described in the text. Circles show the corrected energy differences at the corrected temperatures for the demons (open circles) and spins (solid circles near the center of the open circles). The dot-dashed line shows zero deviation from the expected energies.

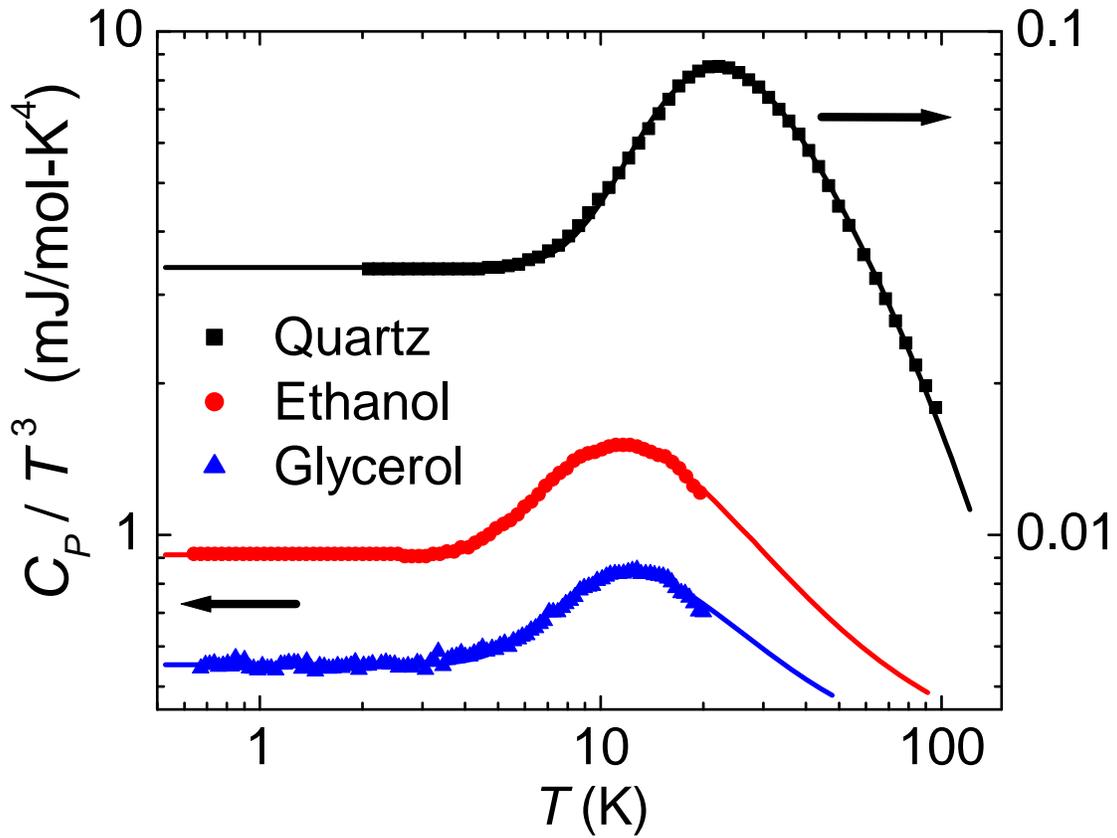

FIG. 4. (Color online) Normalized specific heat as a function of temperature on a log-log plot. Symbols come from the measured specific heat [1,3] of three different crystalline substances: quartz, ethanol, and glycerol (from top to bottom). Debye theory predicts constant $C_P/T^3$ at these low temperatures, as seen below 3 K. Solid lines come from best fits to the excess specific heat that peaks above 10 K using Eq. (3) as described in the text. Agreement with the data suggests that the excess specific heat comes from an effective interaction between units of energy that are adsorbed on the lattice.

| Localization length $L$ (sites) | Energy-transfer amplitude $A$ | Energy-quantum activity $a = \exp(\mu/k_B T_{RW})$ | Effective potential $-w/\varepsilon$ |
|---|---|---|---|
| 3 | 0.322±0.012 | 0.629±0.013 | 3.90±0.06 |
| 4 | 0.132±0.003 | 0.439±0.008 | 4.03±0.04 |
| 6 | 0.113±0.013 | 0.480±0.04 | 4.02±0.16 |
| 8 | 0.071±0.011 | 0.419±0.04 | 4.05±0.19 |
| Substance | Excess specific heat constant $C$ (mJ/mol-$K^4$) | $a = \exp(\mu/k_B T)$ | |
| Quartz | 0.0197±0.005 | 0.503±0.014 | 2.936±0.015 |
| Ethanol | 0.22±0.02 | 0.415±0.04 | 2.79±0.14 |
| Glycerol | 0.11±0.03 | 0.430±0.11 | 3.0±0.8 |

**Table 1.** Upper rows: Fit parameters for the energy-quanta adsorption model from IL dynamics where excited demons reflect on the length scale given. The model characterizes the energy offset from the BE distribution for demons $n = n_0 + A/\xi$, and from the FD distribution for spins $m = m_0 - A/\xi$, where $\xi = 1 + 2ae^{-\varepsilon/k_B T_{RW}} + a^2 e^{-2\varepsilon/k_B T_{RW}} e^{-w/k_B T_{RW}}$. Note that when obtaining $A$ and $a$, the value of $-w/\varepsilon$ was fixed at 4 to reduce uncertainty. Lower rows: Similar fit parameters from the excess specific heat of imperfect crystals at low temperatures using Eq. (3).